\documentclass[runningheads]{llncs}
\usepackage{subcaption}
\usepackage[T1]{fontenc}
\usepackage{array}
\usepackage{svg}
\usepackage{tikz}
\usepackage{graphicx}
\usepackage{amssymb}
\usepackage{diagbox}
\usepackage{amsmath}
\usepackage{xcolor}
\usepackage{babel}
\usepackage{multirow}
\usepackage{enumitem}
\usepackage{pgfplots}
\usepgfplotslibrary{statistics}
\pgfplotsset{{compat=newest}}
\usetikzlibrary{calc}
\usepackage{pgfplotstable}
\usepackage{hyperref}
\newcommand{\tikzcircle}[2][red,fill=red]{\tikz[baseline=-0.65ex]\draw[#1,radius=#2] (0,0) circle ;}%

\begin{document}
\sloppy
\definecolor{bblue}{HTML}{4F81BD}
\definecolor{rred}{HTML}{C0504D}
\definecolor{ggreen}{HTML}{9BBB59}
\definecolor{ppurple}{HTML}{9F4C7C}
\definecolor{bittersweet}{HTML}{00B3B8}

\author{Malo Alefsen\inst{1} \and
 Eugene Vorontsov\inst{3} \and
Samuel Kadoury\inst{1,2}}
%
\authorrunning{Alefsen et al.}
%
\institute{Ecole Polytechnique de Montréal, Montréal, QC, Canada \and Centre de Recherche du CHUM, Montréal, QC, Canada \and Paige, Montréal, QC, Canada}

\title{M-GenSeg: Domain Adaptation For Target Modality Tumor Segmentation With Annotation-Efficient Supervision}
\titlerunning{M-GenSeg: Domain Adaptation For Target Modality Tumor Segmentation}
%
%
%
%
\maketitle              
\begin{abstract}

    Automated medical image segmentation using deep neural networks typically requires substantial supervised training. However, these models fail to generalize well across different imaging modalities. This shortcoming, amplified by the limited availability of expert annotated data, has been hampering the deployment of such methods at a larger scale across modalities. To address these issues, we propose M-GenSeg, a new semi-supervised generative training strategy for cross-modality tumor segmentation on unpaired bi-modal datasets. With the addition of known healthy images, an unsupervised objective encourages the model to disentangling tumors from the background, which parallels the segmentation task. Then, by teaching the model to convert images across modalities, we leverage available pixel-level annotations from the source modality to enable segmentation in the unannotated target modality. We evaluated the performance on a brain tumor segmentation dataset composed of four different contrast sequences from the public BraTS 2020 challenge data. We report consistent improvement in Dice scores over state-of-the-art domain-adaptive baselines on the unannotated target modality. Unlike the prior art, M-GenSeg also introduces the ability to train with a partially annotated source modality.

\keywords{Image Segmentation  \and Semi-supervised Learning \and Unpaired Image-to-image Translation.}
\end{abstract}

\section{Introduction}

Deep learning methods have demonstrated their tremendous potential when it comes to medical image segmentation. However, the success of most existing architectures relies on the availability of pixel-level annotations, which are difficult to produce \cite{challenges}. Furthermore, these methods are known to be inadequately equipped for distribution shifts. Therefore, cross-modality generalization is needed when one imaging modality has insufficient training data. For instance, conditions such as Vestibular Schwannoma, where new hrT2 sequences are set to replace ceT1 for diagnosis to mitigate the use of contrast agents, is a sample use case \cite{crossmoda}. Recently Billot et al. \cite{synthseg} proposed a domain randomisation strategy to segment images from a wide range of target contrasts without any fine-tuning. The method demonstrated great generalization capability for brain parcellation, but the model performance when exposed to tumors and pathologies was not quantified. This challenge could also be addressed through unsupervised domain-adaptive approaches, which transfer the knowledge available in the "source" modality $S$ from pixel-level labels to the "target" imaging modality $T$ lacking annotations \cite{Domain}.

Several generative models attempt to generalize to a target modality by performing unsupervised domain adaptation through image-to-image translation and image reconstruction. In \cite{disentangle}, by learning to translate between CT and MR cardiac images, the proposed method jointly disentangles the domain specific and domain invariant features between each modality and trains a segmenter from the domain invariant features. Other methods \cite{attent,SynSeg,Synergistic,self,constrained,cycada,x_ray} also integrate this translation approach, but the segmenter is trained in an end-to-end manner on the synthetic target images generated from the source modality using a CycleGAN \cite{cyclegan} model. These methods perform well but do not explicitly use the unannotated target modality data to further improve the segmentation.

\begin{figure}[t]\centering
\includegraphics[width=0.97\columnwidth]{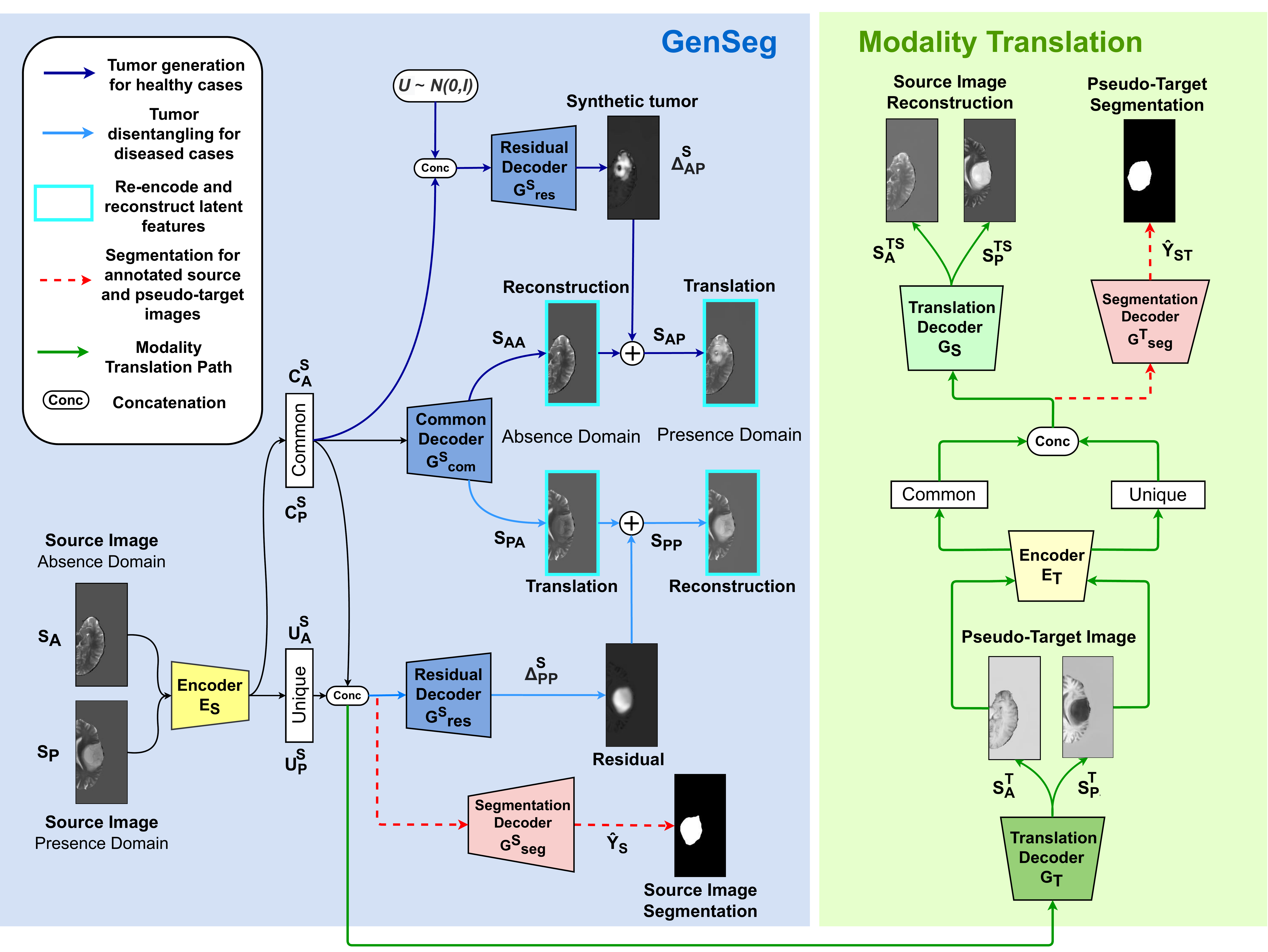}
\caption{M-GenSeg: Latent representations are shared for simultaneous cross-modality translation (green) and semi-supervised segmentation (blue). Source images are passed through the source GenSeg module and the S$\rightarrow$T*$\rightarrow$S modality translation cycle. Domain adaptation is achieved when training the segmentation on annotated pseudo-target T* images ($\mathbf{S^T_P}$). It is not shown but, symmetrically, target images are treated in an other branch to train the T$\rightarrow$S$\rightarrow$T cyclic translation, and the target GenSeg module to further close the domain gap.}
\label{diag1}
\end{figure}

In this paper, we propose M-GenSeg, a novel training strategy for cross-modality domain adaptation, as illustrated in Fig. \ref{diag1}. This work leverages and extends GenSeg \cite{GenSeg}, a generative method that uses image-level "diseased" or "healthy" labels for semi-supervised segmentation. Given these labels, the model imposes an image-to-image translation objective between the image domain presenting tumor lesions and the domain corresponding to an absence of lesions. Therefore, like in low-rank atlas based methods \cite{rank1,rank2,rank3} the model is taught to find and remove a lesion, which acts as a guide for the segmentation. We incorporate cross-modality image segmentation with an image-to-image translation objective between source and target modalities. We hypothesize both objectives are complementary since GenSeg helps localizing the tumors on unannotated target images, while modality translation enables fine-tuning the segmenter on the target modality by displaying annotated pseudo-target images. We evaluate M-GenSeg on a modified version of the BraTS 2020 dataset, in which each type of sequence (T1, T2, T1ce and FLAIR) is considered as a distinct modality. We demonstrate that our model can better generalize than other state-of-the-art methods to the target modality.

\section{Methods}

\subsection{M-GenSeg : semi-supervised segmentation}

\subsubsection{Healthy-diseased translation.}
We propose to integrate image-level supervision to the cross-modality segmentation task with GenSeg, a model that introduces translation between domains with a presence (P) or absence (A) of tumor lesions. Leveraging this framework has a two-fold advantage here. Indeed, \emph{(i)} training a GenSeg module on the source modality makes the model aware of the tumor appearances in the source images even with limited source pixel-level annotations. This helps to preserve tumor structures during the generation of pseudo-target samples (see section \ref{mod_trans}). Furthermore, \emph{(ii)} training a second GenSeg module on the target modality allows to further close the domain gap by extending the segmentation objective to unannotated target data. 
\newline In order to disentangle the information common to A and P, and the information specific to P, we split the latent representation of each image into a common code $\mathbf{c}$ and a unique code $\mathbf{u}$. Essentially, the common code contains information inherent to both domains, which represents organs and other structures, while the unique code stores features like tumor shapes and location. In the two following paragraphs, we explain P$\rightarrow$A and A$\rightarrow$P translations for source images. The same process is applied for target images by replacing $S$ notation with $T$.

\paragraph{Presence to absence translation.}
\label{p to a}
Given an image $\mathbf{S_{P}}$ of modality S in the presence domain P, we use an encoder $E_S$ to compute the latent representation $\mathbf{[}\mathbf{c_{P}^S},\mathbf{u_{P}^S}\mathbf{]}$. A common decoder $G_{com}^S$ takes as input the common code $\mathbf{c_{P}^S}$ and generates a healthy version $\mathbf{S_{PA}}$ of that image by removing the apparent tumor region. Simultaneously, both common and unique codes are used by a residual decoder $G_{res}^S$ to output a residual image $\mathbf{\Delta_{PP}^S}$, which corresponds to the additive change necessary to shift the generated healthy image back to the presence domain. In other words, the residual is the disentangled tumor that can be added to the generated healthy image to create a reconstruction $\mathbf{S_{PP}}$ of the initial diseased image:
\begin{equation}
\mathbf{S_{PA}}=G_{com}^S(\mathbf{c_{P}^S})\,\text{  and  }\,\mathbf{\Delta_{PP}^S}=G_{res}^S(\mathbf{c_{P}^S},\mathbf{u_{P}^S})\,\text{ and }\,\mathbf{S_{PP}}=\mathbf{S_{PA}}+\mathbf{\Delta_{PP}^S}
\end{equation}

\paragraph{Absence to presence translation.}
\label{a to p}
Concomitantly, a similar path is implemented for images in the healthy domain. Given an image $\mathbf{S_{A}}$ of modality $S$ in domain A, we generate a translated version in domain P. To do so, a synthetic tumor $\mathbf{\Delta_{AP}^S}$ is generated by sampling a code from the normal distribution $\mathcal{N}(0,\mathbf{I})$ and replacing the encoded unique code for that image. The reconstruction $\mathbf{S_{AA}}$ of the original image in domain A and the synthetic diseased image $\mathbf{S_{AP}}$ in domain P are computed from the encoded features $\mathbf{[}\mathbf{c_{A}^S},\mathbf{u_{A}^S}\mathbf{]}$ as follows:
\begin{equation}
\mathbf{S_{AA}}=G_{com}^S(\mathbf{c_{A}^S})\quad\text{and}\quad 
\mathbf{S_{AP}}=\mathbf{S_{AA}}+G_{res}^S(\mathbf{c_{A}^S},\mathbf{u}\sim\mathcal{N}(0,\mathbf{I}))
\end{equation}
Like approaches in \cite{tum_aug_1,tum_aug_2,inpainting} we therefore generate diseased samples from healthy ones for data augmentation. However, M-GenSeg aims primarily at tackling cross-modality lesion segmentation tasks, which is not addressed in these studies. Furthermore, note that these methods are limited to data augmentation and do not incorporate any unannotated diseased samples when training the segmentation network, as achieved by our model with the P$\rightarrow$A translation. 

\subsubsection{Modality translation.}
\label{mod_trans}
Our objective is to learn to segment tumor lesions in a target modality by reusing potentially scarce image annotations in a source modality. Note that for each modality $m\in\{S,T\}$, M-GenSeg holds a segmentation decoder $G_{seg}^m$ that shares most of its weights with the residual decoder $G_{res}^m$, but has its own set of normalization parameters and a supplementary classifying layer. Thus, through the Absence and Presence translations, these segmenters have already learned how to disentangle the tumor from the background. However, supervised training on a few example annotations is still required to learn how to transform the resulting residual representation into appropriate segmentation maps. While this is a fairly straightforward task for the source modality using pixel-level annotations, achieving this for the target modality is more complex, justifying the second unsupervised translation objective between source and target modalities. Based on the CycleGan \cite{cyclegan} approach, modality translations are performed via two distinct generators that share their encoder with the GenSeg task. More precisely, combined with the encoder $E_S$ a decoder $G_T$ enables performing S$\rightarrow$T modality translation, while the encoder $E_T$ and a second decoder $G_S$ perform the T$\rightarrow$S modality translation. To maintain the anatomical information, we ensure cycle-consistency by reconstructing the initial images after mapping them back to their original modality. We note $\mathbf{S_d^T}=G_T\circ E_S(\mathbf{S_d})$ and $\mathbf{S_d^{TS}}=G_S\circ E_T(\mathbf{S_d^{T}})$, respectively the translation and reconstruction of $\mathbf{S_d}$ in the S$\rightarrow$T$\rightarrow$S translation loop, with domain $\mathbf{d}\in\{A,P\}$ and $\circ$ the composition operation. Similarly we have $\mathbf{T_d^S}=G_S\circ E_T(\mathbf{T_d})$ and $\mathbf{T_d^{ST}}=G_T\circ E_S(\mathbf{T_d^{S}})$ for the T$\rightarrow$S$\rightarrow$T cycle.
\newline
Note that to perform the domain adaptation, training the model to segment only the pseudo-target images generated by the S$\rightarrow$T modality generator would suffice (in addition to the diseased/healthy target translation). However, training the segmentation on diseased source images also imposes additional constraints on encoder $E_S$, ensuring the preservation of tumor structures. This constraint proves beneficial for the translation decoder $G_T$ as it generates pseudo-target tumoral samples that are more reliable. 
Segmentation is therefore trained on both diseased source images $\mathbf{S_P}$ and their corresponding synthetic target images $\mathbf{S_P^T}$, when provided with annotations $\mathbf{y_S}$. To such an extent, two segmentation masks are predicted 
$\mathbf{\hat{y}_S}=G_{seg}^S\circ E_S(\mathbf{S_P})$ and $\mathbf{\hat{y}_{ST}}=G_{seg}^T\circ E_T(\mathbf{S^T_P})$.

\subsection{Loss functions}

\subsubsection{Segmentation Loss.}
For the segmentation objective, we compute a soft Dice loss \cite{vor} on the predictions for both labelled source images and their translations:\begin{equation} \mathcal{L}_{seg}=Dice\left(\mathbf{{y}_{S}},\mathbf{\hat{y}_S}\right)+Dice\left(\mathbf{y_S},\mathbf{\hat{y}_{ST}}\right)
\end{equation}
\textbf{Reconstruction Losses.}
$\mathcal{L}_{cyc}^{mod}$ and $\mathcal{L}_{rec}^{Gen}$ respectively impose pixel-level image reconstruction constraints on modality translation and GenSeg tasks. Note that $\mathcal{L}_{1}$ refers to the standard L1 norm:
\begin{equation}
\begin{array}{c}
\mathcal{L}_{cyc}^{mod}=\mathcal{L}_{1}\left(\mathbf{S_{A}^{TS}},\mathbf{S_{A}}\right)+\mathcal{L}_{1}\left(\mathbf{T_{A}^{ST}},\mathbf{T_{A}}\right)+\mathcal{L}_{1}\left(\mathbf{S_{P}^{TS}},\mathbf{S_{P}}\right)+\mathcal{L}_{1}\left(\mathbf{T_{P}^{ST}},\mathbf{T_{P}}\right)\\[\medskipamount]
\mathcal{L}_{rec}^{Gen}=\mathcal{L}_{1}\left(\mathbf{S_{AA}},\mathbf{S_{A}}\right)+\mathcal{L}_{1}\left(\mathbf{S_{PP}},\mathbf{S_{P}}\right)+\mathcal{L}_{1}\left(\mathbf{T_{AA}},\mathbf{T_{A}}\right)+\mathcal{L}_{1}\left(\mathbf{T_{PP}},\mathbf{T_{P}}\right)
\end{array}
\end{equation}

\noindent Moreover, like in \cite{GenSeg} we compute a loss $\mathcal{L}_{lat}^{Gen}$ that ensures that the translation task holds the information relative to the initial image, by reconstructing their latent codes with the L1 norm. It also enforces the distribution of unique codes to match the prior $\mathcal{N}(0,\mathbf{I})$ by making $\mathbf{u_{AP}}$ match $\mathbf{u}$, where $\mathbf{u_{AP}}$ is obtained by encoding the fake diseased sample $\mathbf{x_{AP}}$ produced with random sample $\mathbf{u}$.\vspace{0.2cm}
\newline 
\textbf{Adversarial Loss.}
 For the healthy-diseased translation adversarial objective, we compute a hinge loss $\mathcal{L}_{adv}^{Gen}$ as in GenSeg, learning to discriminate between pairs of real/synthetic images of the same output domain and always in the same imaging modality, e.g. $\mathbf{S_A}$ vs $\mathbf{{S}_{PA}}$. In the modality translation task, the $\mathcal{L}_{adv}^{mod}$ loss is computed between pairs of images of the same modality without distinction between domains $A$ and $P$, e.g. $\{\mathbf{S_A},\mathbf{S_P}\}$ vs $\{\mathbf{{T}^S_{A}},\mathbf{{T}^S_{P}}\}$ \vspace{0.2cm}
\newline
\textbf{Overall Loss.}
The overall loss for M-GenSeg is a weighted sum of the aforementioned losses. These are tuned separately. All weights sum to 1. First, $\lambda_{adv}^{Gen}$, $\lambda_{rec}^{Gen}$, and $\lambda_{lat}^{Gen}$ weights are tuned for successful translation between diseased and healthy images. Then, $\lambda_{adv}^{mod}$ and $\lambda_{cyc}^{mod}$ are tuned for successful modality translation. Finally, $\lambda_{seg}$ is tuned for segmentation performance.

\begin{equation}
\begin{array}{cc}
\mathcal{L}_{Total}&=\lambda_{seg}\mathcal{L}_{seg}+ \lambda_{adv}^{mod}\mathcal{L}_{adv}^{mod}+\lambda_{cyc}^{mod}\mathcal{L}_{cyc}^{mod}\label{lambda}\\[\medskipamount]
&+\lambda_{adv}^{Gen}\mathcal{L}_{adv}^{Gen}+\lambda_{rec}^{Gen}\mathcal{L}_{rec}^{Gen}+\lambda_{lat}^{Gen}\mathcal{L}_{lat}^{Gen}
\end{array}
\end{equation}

\subsection{Implementation Details}

\begin{figure}[t]
	\begin{subfigure}[c]{0.35\linewidth}
		\centering
		\includegraphics[width=\textwidth]{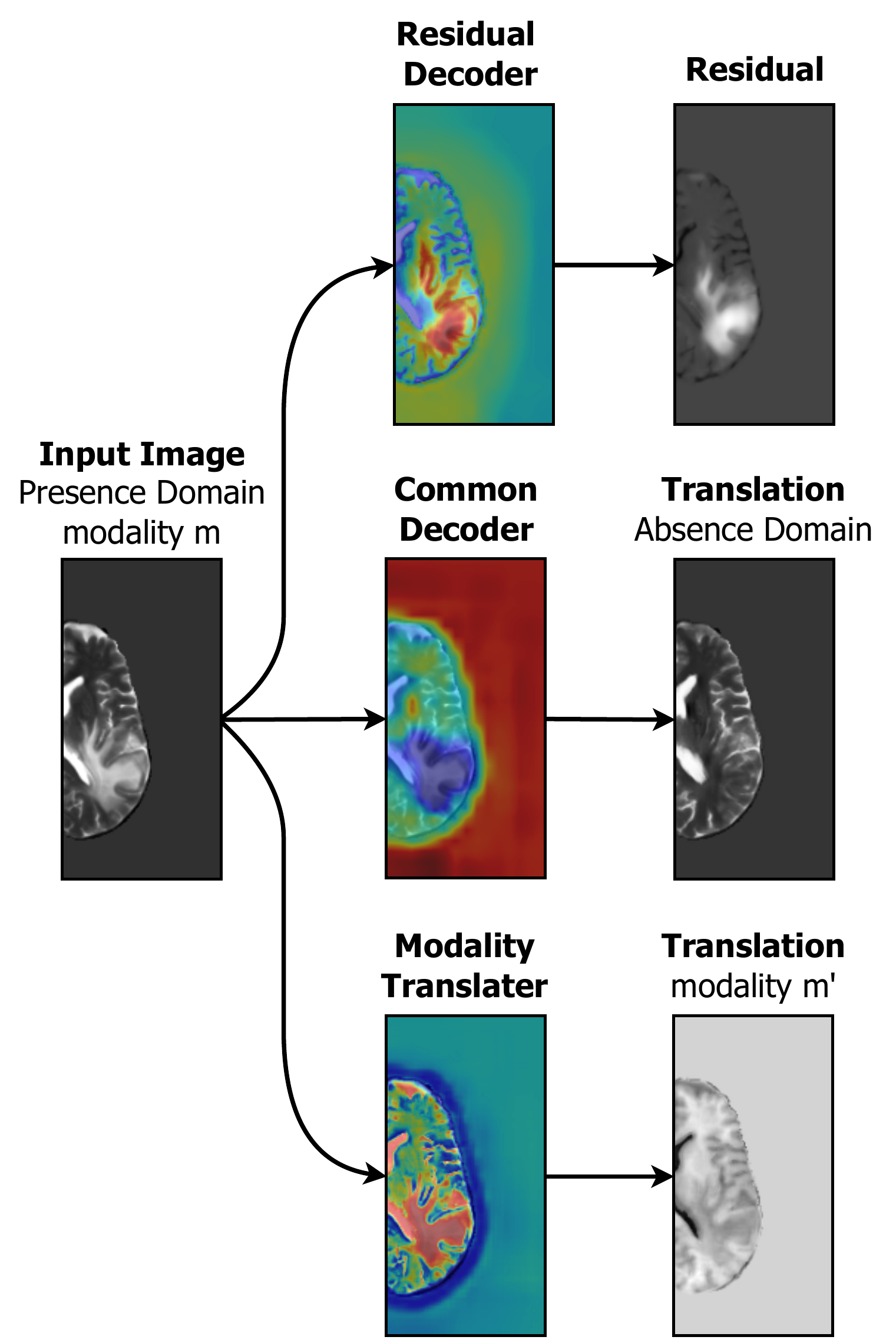}
  \vspace{0.005cm}
  \caption{}
  \label{attention}
	\end{subfigure}\hfill
	\begin{subfigure}[c]{0.48\linewidth}
		\centering
		\includegraphics[width=\textwidth]{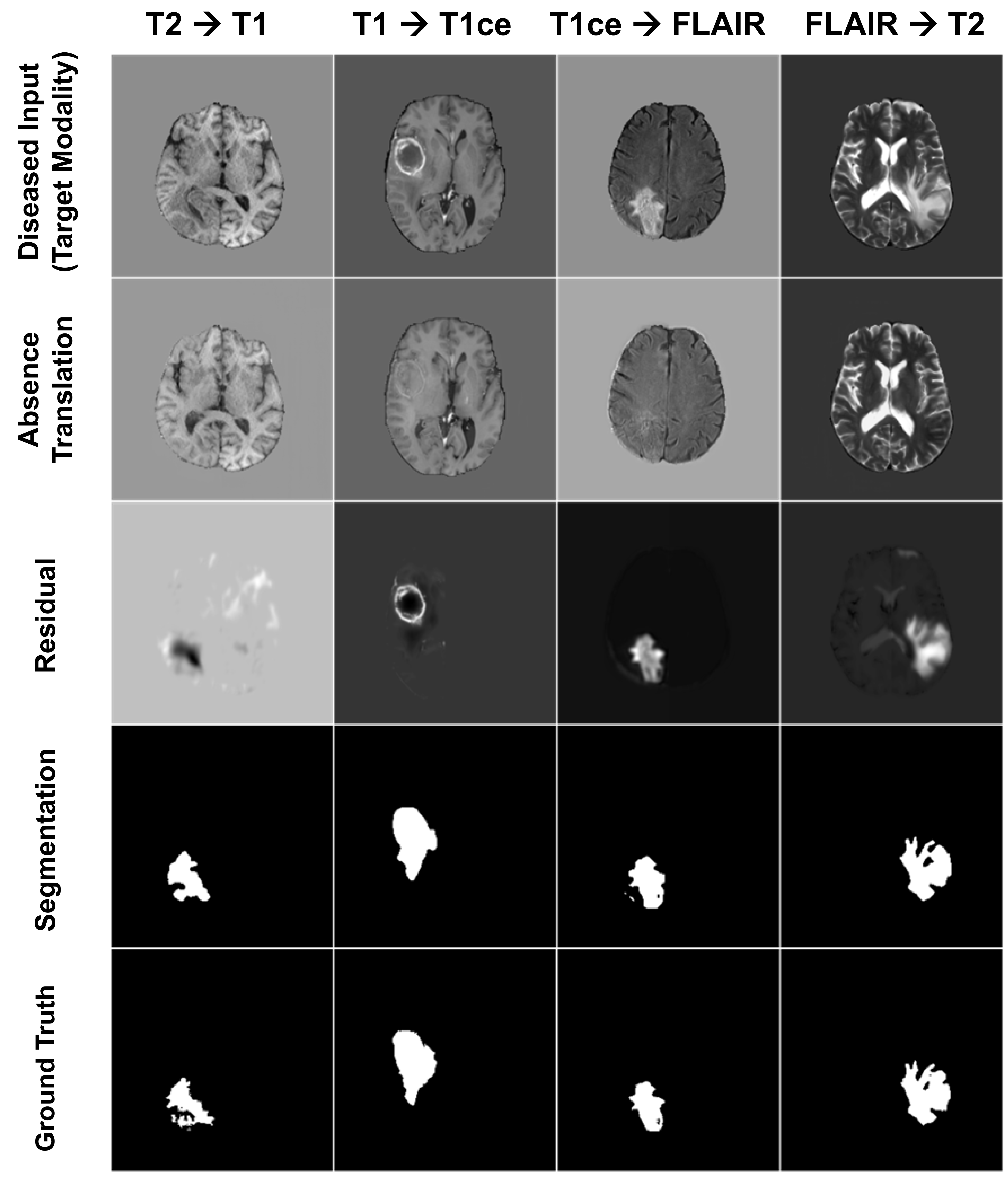}
  \caption{}
  \label{mbrats}
	\end{subfigure}
\caption{(a) Attention maps for Presence $\rightarrow$ Absence and modality translations. Red indicates areas of focus while dark blue correspond to locations ignored by the network. (b) Examples of translations from Presence to Absence domains and resulting segmentation. Each column represents a domain adaptation scenario where target modality had no pixel-level annotations provided.}	
\end{figure}

\textbf{Training and hyper-parameters.}
All models are implemented using PyTorch and are trained on one NVIDIA A100 GPU with 40 GB memory. We used a batch size of 15, an AMSGrad optimizer ($\beta_1=0.5$ and $\beta_2=0.999$) and a learning rate of $10^{-4}$. Our models were trained for 300 epochs and weights of the segmentation model with the highest validation Dice score were saved for evaluation. The same on-the-fly data augmentation as in \cite{GenSeg} was applied for all runs. Each training experiment was repeated three times with a different random seed for weight initialization. The performance reported is the mean of all test Dice scores, with standard deviation, across the three runs. The following parameters yielded both great modality and absence/presence translations : $\lambda_{adv}^{mod}=3$, $\lambda_{cyc}^{mod}=20$, $\lambda_{adv}^{Gen}=6$, $\lambda_{rec}^{Gen}=20$ and $\lambda_{lat}^{Gen}=2$. Note that optimal $\lambda_{seg}$ varies depending on the fraction of pixel-level annotations provided to the network for training.
\vspace{0.2cm}
\newline
\textbf{Architecture.}
One distinct encoder, common decoder, residual/segmentation decoder, and modality translation decoder are used for each modality. The architecture used for encoders, decoders and discriminators is the same as in \cite{GenSeg}. However, in order to give insight on the model's behaviour and properly choose the semantic information relevant for each objective, we introduced attention gates \cite{attentionunet} in the skip connections. Fig. \ref{attention} shows the attention maps generated for each type of decoder. As expected, residual decoders focus towards tumor areas. More interestingly, in order not to disturb the process of healthy image generation, common decoders avoid lesion locations. Finally, modality translators tend to focus on salient details of the brain tissue, which facilitates contrast redefinition needed for accurate translation.

\section{Experimental Results}

\subsection{Datasets}

Experiments were performed on the BraTS 2020 challenge dataset \cite{b1,b2,b3}, adapted for the cross-modality tumor segmentation problem where images are known to be diseased or healthy. Amongst the 369 brain volumes available in BraTS, 37 were allocated each for validation and test steps, while the 295 left were used for training. 
We split the 3D brain volumes into 2 hemispheres and extracted 2D axial slices. Any slices with at least 1\% tumor by brain surface area were considered diseased. Those that didn't show any tumor lesion were labelled as healthy images. Datasets were then assembled from each distinct pair of the four MRI contrasts available (T1, T2, T1ce and FLAIR). To constitute unpaired training data, we used only one modality (source or target) per training volume. All the images are provided with healthy/diseased weak labels, distinct from the pixel-level annotations that we provide only to a subset of the data. Note that the interest for cross-sequence segmentation is limited if multi-parametric acquisitions are performed as is the case in BraTS. However, this modified version of the dataset provides an excellent study case for the evaluation of any modality adaptation method for tumor segmentation. 

\subsection{Model Evaluation}

\begin{figure}[t]\centering
\begin{tikzpicture} 
    \begin{axis}[%
    hide axis,
    xmin=10,
    xmax=50,
    ymin=0,
    ymax=0.2,
    legend cell align=left,
        legend image code/.code={%
                    \draw[#1, draw=black] (0.025cm,-0.04cm) rectangle (0.3cm,0.07cm);
                },
        legend style={legend columns=5, anchor=south west,              nodes={scale=0.7, transform shape},
                at={(0.5,1.03)},
                column sep=1ex
        },
        legend image post style={scale=2}
    ]
    \addlegendimage{black, fill=purple!70, mark=none}
    \addlegendentry{No adaptation};
    \addlegendimage{black,fill=blue!40, mark=none};
    \addlegendentry{AttENT};
    \addlegendimage{black,fill=cyan!40, mark=none}
    \addlegendentry{AccSegNet};
    \addlegendimage{black,fill=green!40, mark=none};
    \addlegendentry{M-GenSeg};
    \addlegendimage{black,fill=orange!40, mark=none}
    \addlegendentry{UAGAN};
    \end{axis}
\end{tikzpicture}
\vfill
\centering
{\resizebox{0.24\textwidth}{!}{\begin{tikzpicture}
    \begin{axis}[
        label style = {font=\large},
        ticklabel style = {font=\large},
        yticklabel style = {font=\small},
        ylabel={T1 Dice},        
        width  = 0.5\textwidth,
        height = 5cm,
        ymax=1.05,
        grid style=dashed, 
        major x tick style = transparent,
        ybar=0.1*\pgflinewidth,
        bar width=6.8pt,
        ymajorgrids = true,
        grid style=dashed,    
        xlabel={Target : T1},
        ytick={0,0.2,0.4,0.6,0.8,1},
        yticklabels={0, 0.2, 0.4, 0.6, 0.8, 1},
        symbolic x coords={T1ce, FLAIR, T2},
        symbolic x coords={T1ce, FLAIR, T2},
        xtick = data,
        scaled y ticks = false,
        enlarge x limits=0.3,
        ymin=0,
        legend cell align=left,
        legend image code/.code={%
                    \draw[#1, draw=black] (0.05cm,-0.08cm) rectangle (0.6cm,0.1cm);1
                },
        legend style={legend columns=5, anchor=south west,              nodes={scale=1, transform shape},
                at={(0.5,1.03)},
                column sep=1ex
        },
        legend image post style={scale=2}
    ]
        \addplot+[style={black,fill=purple!70,mark=none}, error bars/.cd, y dir=both, y explicit, error bar style=black]
            coordinates {(T1ce, 0.647) +- (0, 0.004)  (FLAIR, 0.140) +- (0, 0.007)  (T2,0.145) +- (0, 0.005) };
            
        \addplot[style={black,fill=blue!40,mark=none}, error bars/.cd, y dir=both, y explicit, error bar style=black]
            coordinates {(T1ce, 0.680) +- (0, 0.013)  (FLAIR, 0.544) +- (0, 0.008)  (T2,0.657) +- (0, 0.003) };
            
        \addplot[style={black,fill=cyan!40,mark=none}, error bars/.cd, y dir=both, y explicit, error bar style=black]
            coordinates {(T1ce, 0.749) +- (0,0.006)  (FLAIR, 0.598) +- (0, 0.012)  (T2,0.664) +- (0, 0.012) };
            
        \addplot[style={black,fill=green!40,mark=none}, error bars/.cd, y dir=both, y explicit, error bar style=black]
            coordinates {(T1ce, 0.766) +- (0, 0.006)  (FLAIR, 0.665) +- (0, 0.007)  (T2,0.694) +- (0, 0.011) };

        \addplot[style={black,fill=orange!40,mark=none}, error bars/.cd, y dir=both, y explicit, error bar style=black]
            coordinates {(T1ce, 0.8) +- (0, 0.007)  (FLAIR, 0.791) +- (0, 0.002)  (T2, 0.782) +- (0, 0.003) };
    \end{axis}
\end{tikzpicture}}}
\hfill
{\resizebox{0.24\textwidth}{!}{\begin{tikzpicture}
    \begin{axis}[
        label style = {font=\Large},
        ticklabel style = {font=\large},
        yticklabel style = {font=\small},
        ylabel={T1ce Dice},        
        width  = 0.5\textwidth,
        height = 5cm,
        ymax=1.05,
        grid style=dashed, 
        major x tick style = transparent,
        ybar=0.1*\pgflinewidth,
        bar width=6.8pt,
        ymajorgrids = true,
        grid style=dashed, 
        xlabel={Target : T1ce},
        ytick={0,0.2,0.4,0.6,0.8,1},
        yticklabels={0, 0.2, 0.4, 0.6, 0.8, 1},
        symbolic x coords={T1, FLAIR, T2},
        xtick = data,
        scaled y ticks = false,
        enlarge x limits=0.3,
        ymin=0,
        legend cell align=left,
        legend style={                nodes={scale=0.5, transform shape},
                at={(0.99,0.99)},
                column sep=1ex
        },
        legend image post style={scale=0.5}
    ]
        \addplot+[style={black,fill=purple!70,mark=none}, error bars/.cd, y dir=both, y explicit, error bar style=black]
            coordinates {(T1, 0.633) +- (0, 0.008)  (FLAIR, 0.261) +- (0, 0.013)  (T2,0.172) +- (0, 0.010) };
            
        \addplot[style={black,fill=blue!40,mark=none}, error bars/.cd, y dir=both, y explicit, error bar style=black]
            coordinates {(T1, 0.687) +- (0, 0.008)  (FLAIR, 0.644) +- (0, 0.003)  (T2,0.619) +- (0, 0.008) };
            
        \addplot[style={black,fill=cyan!40,mark=none}, error bars/.cd, y dir=both, y explicit, error bar style=black]
            coordinates {(T1, 0.727) +- (0,0.009)  (FLAIR, 0.569) +- (0, 0.004)  (T2,0.590) +- (0, 0.014) };
            
        \addplot[style={black,fill=green!40,mark=none}, error bars/.cd, y dir=both, y explicit, error bar style=black]
            coordinates {(T1, 0.753) +- (0, 0.008)  (FLAIR, 0.690) +- (0, 0.011)  (T2,0.698) +- (0, 0.012) };   

        \addplot[style={black,fill=orange!40,mark=none}, error bars/.cd, y dir=both, y explicit, error bar style=black]
            coordinates {(T1, 0.83) +- (0, 0.004)  (FLAIR, 0.812) +- (0, 0.007)  (T2, 0.823) +- (0, 0.003) };
        
    \end{axis}
\end{tikzpicture}}}
\hfill
{\resizebox{0.24\textwidth}{!}{\begin{tikzpicture}
    \begin{axis}[
        label style = {font=\Large},
        ticklabel style = {font=\large},
        yticklabel style = {font=\small},
        ylabel={FLAIR Dice},           
        width  = 0.5\textwidth,
        height = 5cm,
        ymax=1.05,
        grid style=dashed, 
        major x tick style = transparent,
        ybar=0.1*\pgflinewidth,
        bar width=6.8pt,
        ymajorgrids = true,
        xlabel={Target : FLAIR},
        ytick={0,0.2,0.4,0.6,0.8,1},
        yticklabels={0, 0.2, 0.4, 0.6, 0.8, 1},
        symbolic x coords={T1, T1ce, T2},
        xtick = data,
        scaled y ticks = false,
        enlarge x limits=0.3,
        ymin=0,
        legend cell align=left,
        legend style={                nodes={scale=0.5, transform shape},
                at={(0.99,0.99)},
                column sep=1ex
        },
        legend image post style={scale=0.5}
    ]
        \addplot+[style={black,fill=purple!70,mark=none}, error bars/.cd, y dir=both, y explicit, error bar style=black]
            coordinates {(T1, 0.300) +- (0, 0.005)  (T1ce, 0.447) +- (0, 0.010)  (T2,0.747) +- (0, 0.009) };
            
        \addplot[style={black,fill=blue!40,mark=none}, error bars/.cd, y dir=both, y explicit, error bar style=black]
            coordinates {(T1, 0.676) +- (0, 0.009)  (T1ce, 0.700) +- (0, 0.008)  (T2,0.755) +- (0, 0.008) };
            
        \addplot[style={black,fill=cyan!40,mark=none}, error bars/.cd, y dir=both, y explicit, error bar style=black]
            coordinates {(T1, 0.764) +- (0,0.010)  (T1ce, 0.749) +- (0, 0.011)  (T2,0.759) +- (0, 0.006) };
            
        \addplot[style={black,fill=green!40,mark=none}, error bars/.cd, y dir=both, y explicit, error bar style=black]
            coordinates {(T1, 0.781) +- (0, 0.007)  (T1ce, 0.784) +- (0, 0.010)  (T2,0.795) +- (0, 0.006) };

        \addplot[style={black,fill=orange!40,mark=none}, error bars/.cd, y dir=both, y explicit, error bar style=black]
            coordinates {(T1, 0.889) +- (0, 0.003)  (T1ce, 0.879) +- (0, 0.006)  (T2, 0.878) +- (0, 0.005) };
            
    \end{axis}
\end{tikzpicture}}}
\hfill
{\resizebox{0.24\textwidth}{!}{\begin{tikzpicture}
    \begin{axis}[
        label style = {font=\Large},
        ticklabel style = {font=\large},
        yticklabel style = {font=\small},
        ylabel={T2 Dice},            
        width  = 0.5\textwidth,
        height = 5cm,
        ymax=1.05,
        grid style=dashed, 
        major x tick style = transparent,
        ybar=0.1*\pgflinewidth,
        bar width=6.8pt,
        ymajorgrids = true,
        xlabel={Target : T2},
        ytick={0,0.2,0.4,0.6,0.8,1},
        yticklabels={0, 0.2, 0.4, 0.6, 0.8, 1},
        symbolic x coords={T1, T1ce, FLAIR},
        xtick = data,
        scaled y ticks = false,
        enlarge x limits=0.3,
        ymin=0,
        legend cell align=left,
        legend style={                nodes={scale=0.5, transform shape},
                at={(0.99,0.99)},
                column sep=1ex
        },
        legend image post style={scale=0.5}
    ]
        \addplot+[style={black,fill=purple!70,mark=none}, error bars/.cd, y dir=both, y explicit, error bar style=black]
            coordinates {(T1, 0.285) +- (0, 0.006)  (T1ce, 0.363) +- (0, 0.009)  (FLAIR,0.753) +- (0, 0.004) };
            
        \addplot[style={black,fill=blue!40,mark=none}, error bars/.cd, y dir=both, y explicit, error bar style=black]
            coordinates {(T1, 0.781) +- (0, 0.011)  (T1ce, 0.709) +- (0, 0.005)  (FLAIR,0.767) +- (0, 0.006) };
            
        \addplot[style={black,fill=cyan!40,mark=none}, error bars/.cd, y dir=both, y explicit, error bar style=black]
            coordinates {(T1, 0.804) +- (0,0.006)  (T1ce, 0.813) +- (0, 0.011)  (FLAIR,0.810) +- (0, 0.010) };
            
        \addplot[style={black,fill=green!40,mark=none}, error bars/.cd, y dir=both, y explicit, error bar style=black]
            coordinates {(T1, 0.836) +- (0, 0.004)  (T1ce, 0.831) +- (0, 0.009)  (FLAIR,0.840) +- (0, 0.003) };       

        \addplot[style={black,fill=orange!40,mark=none}, error bars/.cd, y dir=both, y explicit, error bar style=black]
            coordinates {(T1, 0.875) +- (0, 0.006)  (T1ce, 0.873) +- (0, 0.008)  (FLAIR, 0.868) +- (0, 0.003) };
    \end{axis}
\end{tikzpicture}}}
\caption{Dice performance on the target modality for each possible source modality. We compare results for M-GenSeg with AccSegNet and AttENT baselines. For reference we also show Dice scores for source supervised segmentation (No adaptation) and UAGAN trained with all source and target annotations.}
\label{raw dice}
\end{figure}
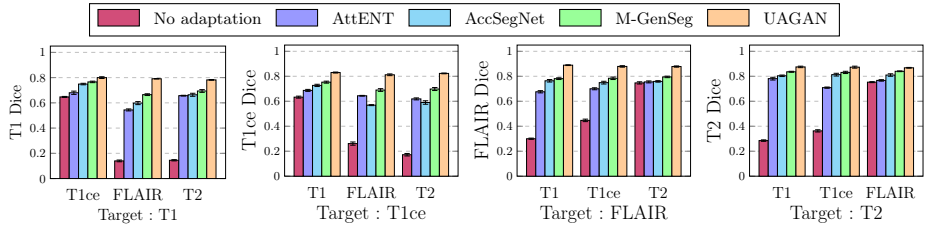

\textbf{Domain adaptation.}
We compared M-GenSeg with AccSegNet \cite{constrained} and AttENT \cite{attent}, two high performance models for domain-adaptative medical image segmentation. To that extent, we performed domain-adaptation experiments with source and target modalities drawn from T1, T2, FLAIR and T1ce. We used available GitHub code for the two baselines and performed fine-tuning on our data. For each possible source/target pair, pixel-level annotations were only retained for the source modality. We show in Fig. \ref{mbrats} several presence to absence translations and segmentation examples on different target modality images. Although no pixel-level annotations were provided for the target modality, tumors were well disentangled from the brain, resulting in a successful presence to absence translation, as well as segmentation. Note that for hypo-intense lesions (T1 and T1ce), M-GenSeg still manages to convert complex residuals into consistent segmentation maps.
We plot in Fig. \ref{raw dice} the Dice performance on the target modality for \emph{(i)} supervised segmentation on source data without domain adaptation, \emph{(ii)} domain adaptation methods and \emph{(iii)} UAGAN \cite{uagan}, a model designed for unpaired multi-modal datasets, trained on all source and target data. Over all modality pairs our model shows an absolute Dice score increase of 0.04 and 0.08, respectively, compared to AccSegNet and AttENT.

\noindent
\begin{minipage}[c]{0.62\textwidth}
\subsubsection{Annotation deficit.} 
\vspace{0.2cm}
M-GenSeg introduces the ability to train with limited pixel-level annotations available in the source modality. We show in Fig. \ref{semi} the Dice scores for models trained when only 1\%, 10\%, 40\%, or 70\% of the source T1 modality and 0\% of the T2 target modality annotations were available. While performance is severely dropping at 1\% of annotations for the baselines, our model shows in comparison only a slight decrease. We thus claim that M-GenSeg can yield robust performance even when a small fraction of the source images is annotated.
\end{minipage}
\hfill
\vspace{0.2cm}
\begin{minipage}[c]{0.333\textwidth}
\centering\resizebox{1\textwidth}{!}{\begin{tikzpicture}
\begin{axis}[
    title style = {font=\huge},
    label style = {font=\Large},
    ticklabel style = {font=\Large},
    xlabel={T1 annotations},
    ylabel={T2 Dice Score},
    ymin=0.1, ymax=0.9,
    ytick={0.1,0.2,0.3,0.4,  0.5,  0.6,  0.7,  0.8},
    xtick={0.01, 0.2, 0.4, 0.6, 0.8, 1},
    xticklabels={1\%, 20\%, 40\%, 60\%, 80\%, 100\%},
    legend style={legend columns=2, cells={align=center},at={(0.5,1.02)},
     anchor=south},
    ymajorgrids=true,
    grid style=dashed
]
        \addplot[style={bblue,mark=diamond*, mark options={scale=2,fill=white}}]
            coordinates {(0.01, 0.734) (0.1, 0.778) (0.4,0.823) (0.7,0.829) (1,0.836)};

        \addplot[style={ggreen,mark=triangle*, mark options={scale=2,fill=white}}]
             coordinates {(0.01, 0.423) (0.1,0.736) (0.4,0.782) (0.7,0.796) (1,0.804)};

        \addplot[style={ppurple,mark=*, mark options={scale=1.5,fill=white}}]
             coordinates {(0.01, 0.478) (0.1,0.674) (0.4,0.721) (0.7,0.747) (1,0.781)};

        \addplot[style={bittersweet,mark=star, mark options={scale=1.5,fill=white}}]
             coordinates {(0.01, 0.353) (0.1,0.251) (0.4,0.218) (0.7,0.198) (1,0.21)};
         \legend{M-GenSeg,AccSegNet,AttENT, Supervised Unet}             
\end{axis}
\end{tikzpicture}\hspace{0.3cm}}
\captionsetup{font=small}

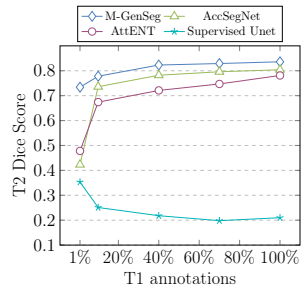
\captionof{figure}{T2 domain adaptation with T1 annotation deficit.}
\label{semi}
\end{minipage}

\noindent\textbf{Reaching supervised performance.}
We report that, when the target modality is completely unannotated, M-GenSeg reaches 90\% of UAGAN's performance (vs $81\%$ and $85\%$ for AttENT and AccSegNet). Further experiments showed that with a fully annotated source modality, it is sufficient to annotate 25\% of the target modality to reach 99\% of the performance of fully-supervised UAGAN (e.g. M-GenSeg : $0.861\pm 0.004$ vs UAGAN : $0.872\pm 0.003$ for T1 $\rightarrow$ T2 experiment). Thus, the annotation burden could be reduced with M-GenSeg.

\subsection{Ablation Experiments}
We conducted ablation tests to validate our methodological choices. We report in Table \ref{ablation} the relative loss in Dice scores on target modality as compared to the proposed model. 
We assessed the value of doing image-level supervision by setting all the $\lambda^{Gen}$ loss weights to 0 \tikzcircle[fill=cyan!40]{3pt}. Also, we showed that training modality translation only on diseased data is sufficient \tikzcircle[fill=blue!40]{3pt}. However, doing it for healthy data as well provides additional training examples for this task. Likewise, performing translation from absence to presence domain is not necessary \tikzcircle[fill=purple!70]{3pt} but makes more efficient use of the data. Finally, we evaluated M-GenSeg with separate latent spaces \tikzcircle[fill=green!40]{3pt} for the image-level supervision and modality translation, and we contend that M-GenSeg efficiently combines both tasks when the latent representations share model updates.

\begin{center}
\parbox{0.62\linewidth}{
\centering
\resizebox{0.62\textwidth}{!}{
\begin{minipage}[c]{\textwidth}
\centering\parbox{\linewidth}{
\centering
\resizebox{\textwidth}{!}{
\begin{tabular}{lcccc}
\firsthline
Ablation & Mean & & Std & \\
\hline
\tikzcircle[fill=cyan!40]{3pt} No image-level supervision      & -8.22 & $\pm$ &  2.71  & \%   \\
\tikzcircle[fill=blue!40]{3pt} No healthy modality translation   & -2.41 & $\pm$ &  1.29    & \%     \\
\tikzcircle[fill=purple!70]{3pt} No absence to presence translation       & -3.84 & $\pm$ &  1.71    & \%     \\
\tikzcircle[fill=green!40]{3pt} Unshared latent spaces       & -4.39 & $\pm$ &  1.91   & \%    \\
\lasthline
\end{tabular}}
\captionsetup{font=Large}
\captionof{table}{Ablation studies : relative Dice change on target modality.}\label{ablation}}
\end{minipage}}}
\end{center}

\section{Conclusion}

We propose M-GenSeg, a new framework for unpaired cross-modality tumor segmentation. We show that M-GenSeg is an annotation-efficient framework that greatly reduces the performance gap due to domain shift in cross-modality tumor segmentation. We claim that healthy tissues, if adequately incorporated to the training process of neural networks like in M-GenSeg, can help to better delineate tumor lesions in segmentation tasks. However, top performing methods on BraTS are 3D models. Thus, future work will explore the use of full 3D images rather than 2D slices, along with more optimal architectures. Our code is available: \url{https://github.com/MaloADBA/MGenSeg_2D}.

\bibliographystyle{IEEEtran}

\bibliography{main}

\end{document}